\documentclass[iop]{emulateapj}

\newcommand{\ps}{Pan-STARRS1}
\newcommand{\pc}{\texttt{pCIGALE}}
\usepackage{url}

\begin{document}
\title{The host galaxy properties of variability selected AGN in the Pan-STARRS1 Medium-Deep Survey}

\author{S. Heinis\altaffilmark{1}, S. Gezari\altaffilmark{1}, S. Kumar\altaffilmark{1}, W. S. Burgett\altaffilmark{2}, H. Flewelling\altaffilmark{2}, M. E. Huber\altaffilmark{2}, N. Kaiser\altaffilmark{2}, R. J. Wainscoat\altaffilmark{2},C. Waters\altaffilmark{2}}


\altaffiltext{1}{Department of Astronomy, University of Maryland, College Park, MD, USA}
\altaffiltext{2}{Institute for Astronomy, University of Hawaii at Manoa, Honolulu, HI 96822, USA}






\begin{abstract}
We study the properties of 975 active galactic nuclei (AGN) selected by variability in the
\ps~Medium-Deep Survey. Using complementary multi wavelength data from
the ultraviolet to the far-infrared, we use SED fitting to determine the AGN and
host properties at $z<1$, and compare to a
well-matched control sample. We confirm the trend previously observed
that the variability amplitude decreases with AGN luminosity, but on
the other hand, we observe that the slope of this relation steepens
with wavelength resulting in a "redder when brighter'' trend at low luminosities. Our results show that AGN are hosted by more massive
hosts than control sample galaxies, while the restframe, dust-corrected
$NUV-r$ color distribution of AGN hosts is similar to control
galaxies. We find a positive correlation between the AGN luminosity
and star formation rate (SFR), independent of redshift. AGN hosts populate the whole range of
SFRs within and outside the Main Sequence of star forming
galaxies. Comparing the distribution of AGN hosts and control
galaxies, we show that AGN hosts are less likely to be hosted by
quiescent galaxies, but more likely to be hosted by Main Sequence or
starburst galaxies.
\end{abstract}


\section{Introduction}

One of the remaining class of puzzling astronomical objects is
active galactic nuclei (AGN). AGN emission is powered by massive black
holes, which are expected to be hosted by most massive galaxies
\citep{Magorrian_1998} and virtually all galaxies. Besides the
interest in high energy physics and strong gravity involved in the processes shaping
their emission, AGN are also crucial in the context of galaxy
formation and evolution. The energy released by AGN in the
interstellar medium has long been invoked by simulation studies to
explain the quenching of star formation activity in massive galaxies
\citep[e.g.][]{Croton_2006}. Evidence for active feedback has also
been brought by observations \citep[e.g.][]{Fabian_2012,Tombesi_2015},
but remains highly controversial. Indeed while a number of studies
found that AGN activity seems to shut down star formation
\citep[e.g.][]{Schawinski_2009, Farrah_2012, Page_2012}, similar
number of studies show that they do not \citep[e.g.][]{Netzer_2009,
  Mullaney_2012a, Rosario_2013}, and that on the other hand AGN
luminosity is positively correlated with star formation rate
\citep{Mullaney_2012b}, while this relation has also been observed to be flat in redshift intervals at $z<2.5$ \citep{Stanley_2015}.\\  Over the last decade, there has
also been considerable work on understanding the triggering mechanisms
of the AGN activity. The most common scenario is that gas-rich galaxy
mergers trigger AGN activity, which in turn quenches star formation
\citep{Hopkins_2008}. Observations do not fully support this scenario
however: while luminous quasar hosts display signatures of current or
past merger activity \citep{Stockton_1982,Canalizo_2001, Bennert_2008}
moderate luminosity AGN on the other hand reside preferentially in
galaxies displaying undisturbed morphologies \citep{Gabor_2009,
  Cisternas_2011, Kocevski_2012}.

Most of the aforementioned studies focused on AGN selected from their
X ray emission or ultraviolet/optical emission lines. While
successful, these studies are biased against heavily obscured AGN
(which can be detected in the far-infrared), and low luminosity AGN,
where the contribution of the host galaxy can be of the same order or
larger than that of the AGN. With the advent of large time domain
surveys such as \ps~ \citep{Kaiser_2010}, and the upcoming LSST
\citep{Ivezic_2008}, a new window is opening for building large
samples of AGN using variability as a complementary selection
\citep[e.g.][]{Sarajedini_2006, Schmidt_2010, Sesar_2007,
  Villforth_2012} albeit with its own set of selection biases,
  including a bias against Type 2 AGN. Indeed, AGN display
variability over the whole spectrum and over a wide range of
timescales, which is thought to be related to accretion disk
instabilities, while long term variability for the so-called
"changing-look quasars'' explained by variable obscuration
\citep[e.g.][]{Cohen_1986, Denney_2014, Tohline_1976, Shappee_2014} or
change in the ionizing flux of the central source itself
\citep{Lamassa_2015} has been observed only in a handful of
objects.\\ Variability selection enables one to probe a large range of
AGN luminosities, and is not biased against low luminosity objects, as
the amplitude of AGN variability actually increases for fainter AGN
\citep[e.g.][]{Hook_1994, Trevese_1994, VandenBerk_2004, Wilhite_2008,
  Bauer_2009, Zuo_2012, GAS_2014}. In this paper, we revisit the
connection between AGN and host galaxy properties, using a sample of
$\sim 1000$ AGN selected by their optical variability in the \ps~
Medium Deep Survey, and complemented by ancillary data from the
ultraviolet (UV) to the far-infrared (FIR). Thanks to this large
wavelength coverage, we are able to separate the AGN and the host
contributions to the observed SED, in order to investigate the link
between host and AGN properties. While large samples of point-like
quasars have previously been built through variability selection
\citep[e.g.][]{MacLeod_2012}, only small sets ($\sim 50$) have been
considered to investigate the AGN-host properties connection
\citep{Villforth_2012, Klesman_2014}.


This paper is organized as follows: in \S \ref{sec_data} we present
our variability selected AGN sample, as well as the ancillary data.
\S \ref{sec_cigale} describes our fitting method to separate the
AGN and host contribution to the obvserved SED. In
\S \ref{sec_results} we present our results, which are further
discussed in \S \ref{sec_discussion}, before concluding in
\S \ref{sec_conclusions}. Throughout we use a $\Lambda-$flat cosmology
($\Omega_{M}=0.3$, $\Omega_{\Lambda}=0.7$, $H_0=70\, \rm{km}.\rm{s}^{-1}.\rm{Mpc}^{-1}$), and a \citet{Chabrier_2003} initial
  mass function (IMF).

\section{Data}\label{sec_data}
\subsection{Variability selected AGN}\label{sec_var_sel_agn}
We use the sample of candidate AGN selected by variability from
\citet{Kumar_2015}. We only recall here the main characteristics of
the method.  We refer the reader to \citet{Kumar_2015} for full
details.  Our sample is based on the classification of extragalactic
variable sources detected during the first 2.5 years of PS1
observations.  During their $\sim 5\,$ month window of
seasonal visibility, each PS1 Medium Deep Field is observed nightly,
cycling through 4 filters (Tonry et al. 2012) ($g_{\rm PS1}
(\lambda_{\rm eff} = 481$ nm), $r_{\rm PS1} (\lambda_{\rm eff} =
617)$ nm), $i_{\rm PS1} (\lambda_{\rm eff} = 752)$ nm), and $z_{\rm
PS1}$ ($\lambda_{\rm eff} = 866)$ nm), with observations in the same
filter every 3 nights, and observations in the $y_{\rm PS1}$
($\lambda_{\rm eff} = 962)$ nm) filter near the full moon, with an
average number of total epochs per filter for this 2.5 year sample of
36.  Nightly images are processed through a frame subtraction analysis
pipeline, and sources are tagged as a transient and published to an
alerts database if they are detected with a signal-to-noise (S/N)
$\ge$ 5 in at least 3 difference images within a time window of 15
days.  While this pipeline was designed to detect supernovae,
variability in the nuclei of galaxies also is detected as positive and
negative excursions in the difference images.  We then match these
transient alerts with a catalog created from the stacked PS1 images
(described in \S 2.2), and only classify the light curves of
transients within the Kron elliptical radius of an extended galaxy
with $i < 24$ mag, the magnitude range for which the star/galaxy
classification is reliable.

The light curves of variable sources in $g_{PS1},
r_{PS1}, i_{PS1},$ and $z_{PS1}$ image differencing are fitted with
five models: a Gamma distribution, a Gaussian distribution, an
analytic supernova (SN) model (all three modeling SN-like
lightcurves); an Ornstein$-$Uhlenbeck (OU) process \citep[modeling
  AGN-like lightcurves,][]{Kelly_2009}; and a constant flux model
(modeling noise). The quality of the fits is then used to classify the
variable sources using a $K$-means clustering algorithm with 3 centers
(SN, AGN, or noise model). Out of 4361 extragalactic transient alerts,
the light curves of 2262 are classified as similar to those of AGN and
considered "nuclear''. We further restrict the sample to objects with
$i_{\rm PS1}>18$, in order to avoid bright galaxies, where the
difference imaging is more likely to introduce image differencing
artifacts that can cause centroid errors and false nuclear
positive. This cut minimizes the contamination of by
non-AGN to 16\%, according to a verification set of objects with
spectroscopic redshifts, while maximizing the number of AGN in the
sample. We are then left with 1768 objects.

\subsection{Pan-STARRS1 data}
We perform our custom reduction of the \ps~Medium Deep data survey.  We use
the stacks generated by the \ps~ image processing pipeline \citep{Magnier_2006}, and also the CFHT $u$ band
data obtained by E. Magnier as follow up of the Medium Deep fields, which covers 65 \% of the survey. We
have at hand 6 bands: $u_{\rm CFHT}$, $g_{\rm PS1}$, $r_{\rm PS1}$, $i_{\rm PS1}$,
$z_{\rm PS1}$, and $y_{\rm PS1}$. We perform photometry using the following
steps; we consider the \ps \textit{skycell} as the smallest entity:
\textit{i)} resample the $u$ band images to the \ps~resolution ($0.25$\arcsec per pixel), and
register all images; \textit{ii)} for each band fit the PSF to a
Moffat function, and match that PSF to the worse PSF in each skycell;
\textit{iii)} using these PSF-matched images, we derive a $\chi^2$
image \citep{Szalay_1999}; \textit{iv)} we perform photometry using
the dual mode of SExtractor \citep{Bertin_1996}, detecting objects in
the $\chi^2$ image and measuring the fluxes in the PSF matched images:
the Kron-like apertures are defined from the $\chi^2$ image and hence
are the same over all bands. The detection threshold, defined by the
$\chi^2$ distribution, is equivalent to a signal-to-noise ratio (SNR) of 1.9$\sigma$.

\subsection{Other datasets}\label{agn_xmatch}
We cross match our AGN sample to a number of other datasets to constrain their SEDs.
\subsubsection{GALEX}
We first cross match our sample with the public GALEX
\citep{Martin_2005}
data\footnote{\url{http://galex.stsci.edu/casjobs/}} using a
5\arcsec~radius using by decreasing order of priority: data from the
Deep Imaging Survey (DIS), the Medium Imaging Survey (MIS), and the
All-sky Imaging Survey (AIS). For the DIS, we use the standard
pipeline data, while for the MIS and the AIS, we use the GCAT Unique
Source Catalogs\footnote{\url{http://archive.stsci.edu/prepds/gcat/}},
which contains the standard pipeline photometry, as well as optimized
photometry for extended objects ($<1\arcmin$).  83\% of sources have a
GALEX cross match. \citet{Budavari_2009} and
  \citet{Seibert_2005} show that 5\arcsec is the optimal search radius
  for GALEX-SDSS matches, and estimate an upper limit of 2\% for the
  number of false matches.

We make then use of the University of Maryland Time Domain Survey data \citep{Gezari_2013, Gezari_2015}. We re-measure the GALEX photometry for objects which are
in images where the exposure time (in FUV or NUV) is greater than the
archive images. We also derive an upper limit at $1\sigma$ for non
detected sources. This process adds photometry for 7\%
sources. Finally, we also attempt to measure the photometry for all
objects without a cross match with GALEX archive data, or derive an
upper limit if no detection is found.

\subsubsection{Spitzer}
We cross match our sample with the Spitzer Enhanced Imaging Products
(SEIP)\footnote{\url{http://irsa.ipac.caltech.edu/data/SPITZER/Enhanced/SEIP/overview.html}}
using a 2\arcsec search radius. 95\% of our sources have a
  match within 1\arcsec. According to the SWIRE release
  2\footnote{\url{http://irsa.ipac.caltech.edu/data/SPITZER/SWIRE/docs/delivery\_doc\_r2\_v2.pdf}},
  we expect only a few percents of false positive matches in that range.

The SEIP contains high
quality photometry in IRAC (3.6, 4.5, 5.8, and 8.$\,\mu $m), and MIPS
(24 $\mu$m) bands. 54\% of objects in our sample have a cross match
with SEIP sources.

\subsubsection{WISE}
We cross match our sample with the custom reduction of the WISE data
from \citet[][unWISE]{Lang_2014b} using a cross match search
  radius of 1\arcsec. \citet{Lang_2014b} performed prior photometry
based on SDSS sources in their own version of the WISE coadds
\citep{Lang_2014a}. This version of the WISE catalog has the advantage
to provide a WISE flux for all SDSS sources, at the SDSS angular
resolution. We find a cross match for 93\% of objects; given
  the fact that unWISE is based on SDSS prior positions, we expect a
  few percents of false positive matches.

\subsubsection{Spectroscopic redshifts}
We cross match our sample with a number of spectroscopic
catalogs,using a search radius of 1\arcsec. We expect only a
  few percents of false matches as all these datasets are based on
  optical data.  We cross match with SDSS DR12 \citep{Alam_2015},
which provides most of the spectroscopic counterparts. We also cross
match with the following surveys: the COSMOS bright spectroscopic
sample \citep{Lilly_2007}, the PRIMUS survey \citep{Coil_2011,
  Cool_2013}, the VIPERS survey \citep{Guzzo_2014}, the VVDS survey
\citep{Lefevre_2004, Lefevre_2005, Garilli_2008}, the DEEP2 survey
\citep{Newman_2013}, and the Veroncat catalogue
\citep{Veron_2010}. Among our 1768 objects, 585 (33\%) have a
spectroscopic redshift, 493 (85\% of the spectroscopic objects) have
been classified as AGN, and 87 (15\%) as galaxies.

\begin{deluxetable*}{cccccccccccccccccc}
  \tablecaption{Statistics of available photometry\label{tab_phot}}
\tablehead{\colhead{ } & \colhead{$FUV$} & \colhead{$NUV$} & \colhead{$u$} & \colhead{$g_{\rm PS1}$} & \colhead{$r_{\rm PS1}$} & \colhead{$i_{\rm PS1}$} & \colhead{$z_{\rm PS1}$} & \colhead{$y_{\rm PS1}$} & \colhead{$i_1$} & \colhead{$i_2$} & \colhead{$i_3$} & \colhead{$i_4$} & \colhead{$m_1$} & \colhead{$W_1$} & \colhead{$W_2$} & \colhead{$W_3$} & \colhead{$W_4$}}
\startdata
 \% detections  & 74 & 86 & 80 & 100 & 100 & 100 & 100 & 100 & 50 & 50 & 45 & 46 & 36 & 93 & 93 & 82 & 67 \\
\% upper limits & 9   & 5   & 0   &    0  & 0     &  0    & 0     & 0     &  0  & 0   & 1   &  1   & 9 & 0    & 0   & 0   & 0
\enddata
\tablenotetext{}{Statistics of available photometry for the AGN sample. For each band, we give the percentage of objects with detections, and the percentage of objects with upper limits. }
\end{deluxetable*}

\subsection{Photometric redshifts}
\begin{figure}
\includegraphics[width=\hsize]{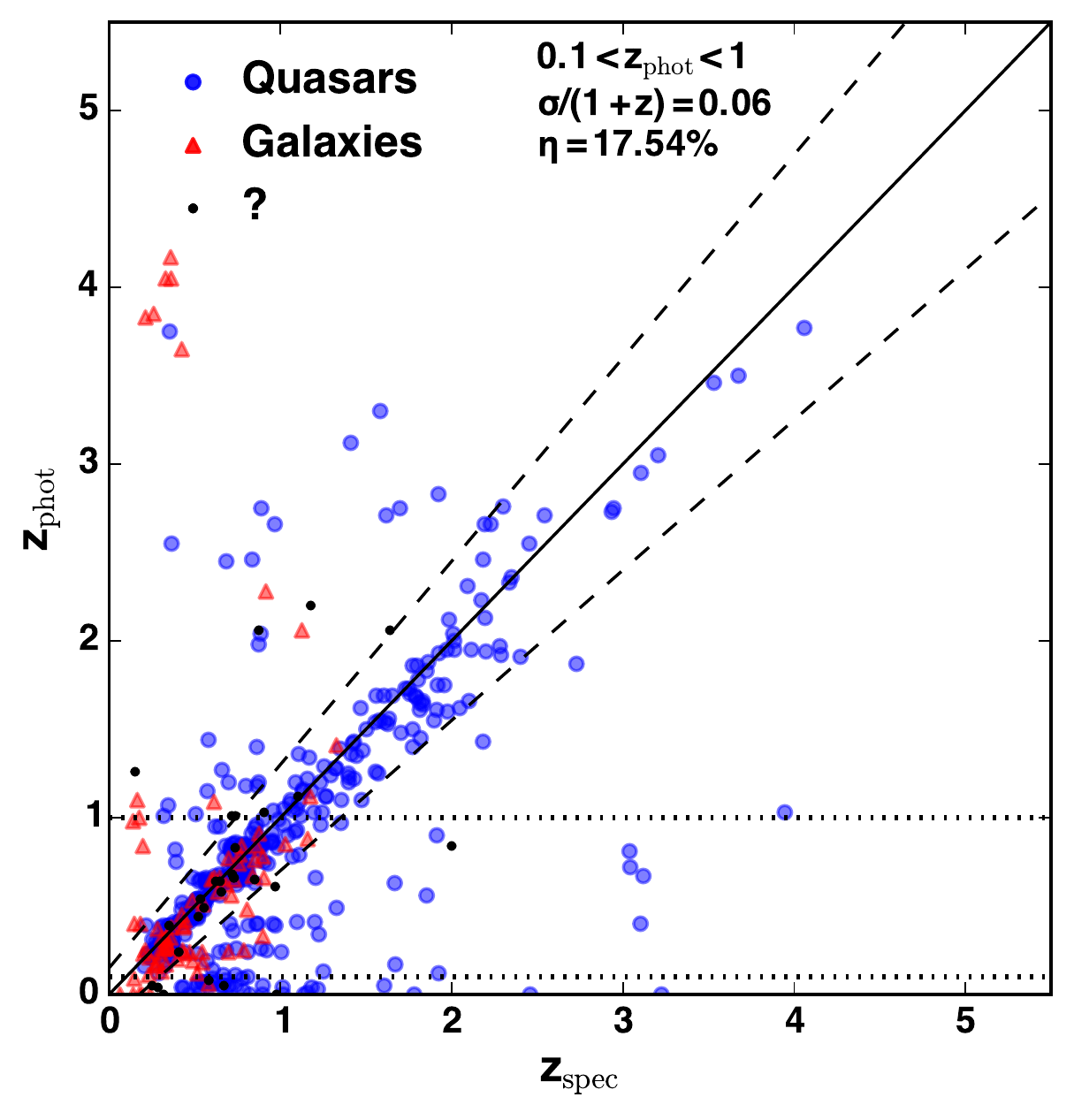}
\caption{Photometric redshifts vs spectroscopic redshifts for the AGN
  sample. Blue circles represent spectroscopic quasars, red triangles
  spectroscopic galaxies, and black dots spectroscopic objects without
  classification available. The solid line shows $z_{\rm spec}=z_{\rm
    phot}$, while the dashed lines correspond to $|z_{\rm spec}-z_{\rm
    phot}|/(1+z_{\rm spec})=0.15$. The dotted lines show our chosen
  cuts in $z_{\rm phot}$. \label{fig_zphot}}
\end{figure}

We estimate photometric redshifts using the code \texttt{lephare}
\citep{Arnouts_1999, Ilbert_2006}. \texttt{lephare} determines
photometric redshifts by fitting the observed photometry to libraries
of galaxies, quasars, and stars template SEDs. We use our training set
built from objects with spectroscopic redshifts to assess the quality
of the photometric redshifts. We do not include here Spitzer
24\,$\mu$m as well as the $W3$ and $W4$ WISE bands, as they probe
ranges of the SED not dominated by stellar emission in the redshift
range we are interested in.  Determining photometric redshifts with
small errors for AGN/quasar dominated objects is notoriously
difficult, as the SEDs of these objects are close to featureless
\citep[see e.g.][]{Richards_2009}, and narrow band photometry is
required to improve significantly the quality of the estimates
\citep{Salvato_2011}. \\We performed extensive tests in order to
obtain the best photometric redshifts for our AGN sample. We first
determine the zero point offsets for the bands. We compared the flux
from best fitting galaxy SED models for spectroscopic galaxies in our
sample to the actual photometry \citep{Ilbert_2009}. In a second step,
we tested a number of combinations for the AGN-dominated SED
templates. We obtain the best results by using a subset of 22
templates out of the 30 templates used by \citet{Salvato_2009}. We
trimmed the list of 30 templates by excluding the templates which are
never retained as best model when the fitting is performed with the
redshift fixed at its spectroscopic value. \\We show a comparison of
the photometric and spectroscopic redshifts on
Fig. \ref{fig_zphot}. Using $\Delta z = z_{\rm spec} - z_{\rm phot}$,
we quantify the error on the photometric redshifts as $err(z)=\Delta
z/(1+z_{\rm spec})$, and use as a global measure the normalized median
absolute deviation $\sigma_{\Delta z/{(1+z_{\rm spec})}} =
1.4826*\textrm{median}(|err(z)-\textrm{median}(err(z))|)$. We consider
as outliers objects with $|err(z)| > 0.15$, and note the percentage of
these objects as $\eta$. For our full spectroscopic sample, the
overall error is $\sigma_{\Delta z/{(1+z_{\rm spec})}} = 0.08$, and
$\eta = 28.1$\%. These numbers are in agreement with those usually
obtained using broadband photometry
\citep[e.g.][]{Salvato_2011}.\\ For this study, we chose hereafter to
restrict the sample to objects with $0.1<z_{\rm phot}<1$; these limits
are shown as dotted lines on Fig. \ref{fig_zphot}. We ensure the upper
limit as the quality of the photometric redshifts decreases
significantly for $z>1$, and the lower limit in order to avoid
outliers a low redshifts. In this range, the error is $\sigma_{\Delta
  z/{(1+z_{\rm spec})}} = 0.06$, and $\eta = 17.5$\%. Our final cuts
leave us with 1160 objects. We further use spectroscopic redshifts
whenever available. Doing so lowers the actual errors on
  photometric redshifts in our sample. In order to derive the
  resulting errors, we assume that the redshift errors for objects
  with spectroscopic redshifts are negligible compared to the
  photometric redshift errors. Assuming objects without spectroscopic
redshifts (757 objects) have the same photometric redshift errors than
those with spectroscopic redshifts (403 objects), the actual error for
our final AGN sample is $\sigma_{\Delta z/{(1+z_{\rm spec})}} = 0.03$,
and the percentage of outliers $\eta=11.5$\%.

\subsection{Control sample}
We build a control sample from the full catalogue from one of the
Medium Deep fields, MD04. We first follow the same procedure as for
the AGN sample described above to obtain photometry in other
wavelengths (see Sect. \ref{agn_xmatch}). We use only extended objects
following our custom star/galaxy separation, based on machine learning
techniques (Heinis et al., submitted). We determine photometric
redshifts using the code \texttt{lephare}. The error on photometric
redshifts is $\sigma_{\Delta z/{(1+z_{\rm spec})}} = 0.05$ and the
outliers percentage $\eta = 11\%$. We keep objects with the same
distribution in $i_{\rm PS1}$ than the AGN sample. We also limit the
sample to $0.1<z_{\rm phot}<1.$, as for the AGN sample. We are left
with 16,401 objects.

\section{SED fitting}\label{sec_cigale}
\subsection{pCIGALE}
We use the code \pc \footnote{\url{http://cigale.lam.fr/}} to perform
SED fitting and estimate physical parameters. \pc~is \texttt{python}
version of the former \texttt{CIGALE} code \citep{Noll_2009}, which
besides the different coding language, also provides new features. \pc~
preserves the same features of \texttt{CIGALE}, however this
new version has been designed for a broader set of scientific
applications as well as improved performance. We provide here a short
description of the latest version of \pc~to date.

\pc~\citep[see also][]{Ciesla_2015} has two different and independent
functions: SED modeling (Boquien et al., in prep.) and SED
fitting (Burgarella et al, in prep.). The SED modeling function
allows one to build a galaxy SED from the UV to the sub-mm based on single
stellar population synthesis models, chosen star formation histories
(SFHs), and energy balance. The full SED is built by re-emitting in
the IR the energy absorbed by dust in the UV-optical.  We use here
delayed star formation histories, which have been shown to reproduce
accurately the SEDs of galaxies over a wide redshift range:
\begin{equation}
  \textrm{SFR}(t) = (t-t_{\rm age})\exp\left(\frac{t-t_{\rm age}}{\tau_{\rm main}}\right)
\end{equation}

We consider the stellar population models of \citet{Bruzual_2003},
which are convolved by the SFH, and then attenuated by dust. We use
the law of \citet{Calzetti_2000} to estimate the extinction by dust,
and the librairies of \citet{Dale_2014} to model the re-emission in
the IR of this energy. Using \texttt {pCIGALE},
  \citet{Buat_2014} showed that the constraint from the IR restframe
  range is essential to derive accurate SFR estimates. In particular
  they found that SFR estimated by SED fitting without IR data are on
  average overestimated by 20\%. In details, low SFR are overestimated
  and large SED are underestimated, by factors up to 2.5. They also
  showed that the intrinsic dispersion in SFR increases by a factor 2
  when no IR data is used.

Finally, \pc~ also allows us to include AGN
emission, to be added to the stellar one, using the templates from
\citet{Fritz_2006}.  The \citet{Fritz_2006} templates consist of two
components: the central source and dust.  The emission of the
central source is assumed to follow power laws with a different index
in three wavelength ranges (spanning $0.001<\lambda<20 \mu$m). The
dust component consists of scattering and the thermal emission from the absorbtion in the UV/optical. The dust torus itself is
modeled using a flared disc geometry \citep{Efstathiou_1995}. The
models of \citet{Fritz_2006} have been extensively tested in previous
work \citep[e.g.][]{Hatziminaoglou_2008, Hatziminaoglou_2010,
  Feltre_2012}. We list in Table \ref{tab_cigale} the parameters we use for the SED
fitting.

\begin{deluxetable*}{ccc}
  \tablecaption{\pc~fitting parameters\label{tab_cigale}}
\tablehead{\colhead{Parameter} & \colhead{Value} & \colhead{Description}}
\startdata
\cutinhead{Delayed Star Formation History}
$\tau_{\rm main}$[Myr] & 10.,50.,100.,500.,1000.,5000.,10000.,100000. & Star formation timescale\\
age [Myr] & 100.,1000.,5000.,7000.,9000.,11000.,13000.,13536. & Age of the oldest stars inthe galaxy \\
\cutinhead{Extinction law \citep{Calzetti_2000}}
$E(B-V)$ & 0.01,0.05,0.1,0.2,0.3,0.5 & Color excess of the stellar continuum\\
\cutinhead{Dust templates \citep{Dale_2014}}
$\alpha_{\rm SF}$  & 1,2,3,4 & Exponent of the intensity of the radiation field\\
\cutinhead{AGN models \citep{Fritz_2006}}
$R_{\rm max}/R_{\rm min}$& 60 & Ratio of the external to internal radius of the dust torus\\
$\tau_{\rm dust}$ & 0.1,0.6,2.0,6.0,10. & Optical depth at 9.7$\mu$m of the dust torus\\
$\beta$ & -0.5& Parameter describing the torus density profile\\
$\gamma$ & 4.& Parameter describing the torus density profile\\
$\Theta$ [deg] & 40.& Opening angle of the dust torus\\
$\psi$ [deg] & 89.99 (Type 1)& Angle between the AGN axis and the line of sight\\
$f_{\rm AGN}$ &0.05,0.1,0.15,0.2,0.25,0.3,0.35,0.4,0.45,0.5,0.55,0.6,0.65,0.7 & AGN fraction in the IR\\
\enddata
\tablenotetext{}{Parameters used for the SED fitting with \pc. For the templates of \citet{Dale_2014}, we fix the AGN fraction to 0.}
\end{deluxetable*}

\pc~creates a library of models combining all parameters; for each
model, a number of properties are derived, such as the stellar mass
($M_*$), the star formation rate averaged over the last 100\,Myr
etc. For the AGN component, we compute for each model the luminosity
at 5100\,\AA~($L_{5100}$), as well as the absolute magnitude in the
$g_{\rm PS1}$ band. \pc~computes the $\chi^2$ statistics for each
model, and builds the probability distribution function (PDF) for each
parameter and derived property using these $\chi^2$ values. The
parameter or property value we use is the average value weighted over
this PDF and the error the standard deviation which thus
  encodes the width of this PDF. In the following, we use only objects
with $\chi^2_{\rm reduced}<5$ (975 objects).

\citet{Ciesla_2015} in particular tested the performance of \pc~to
recover the properties of AGNs in the case of realistic SFHs drawn
from a semi-analytic simulation. \citet{Ciesla_2015} show that \pc~is
able to recover accurately both the properties of the AGNs and the
host galaxy, provided that the observed SED is constrained from the UV
to the FIR restframe, and also that the fraction of the AGN emission
in the IR, $f_{\rm AGN}$ is larger than 0.1.

\subsection{AGN luminosity: SEDs vs spectra}
\begin{figure}
\includegraphics[width=\hsize]{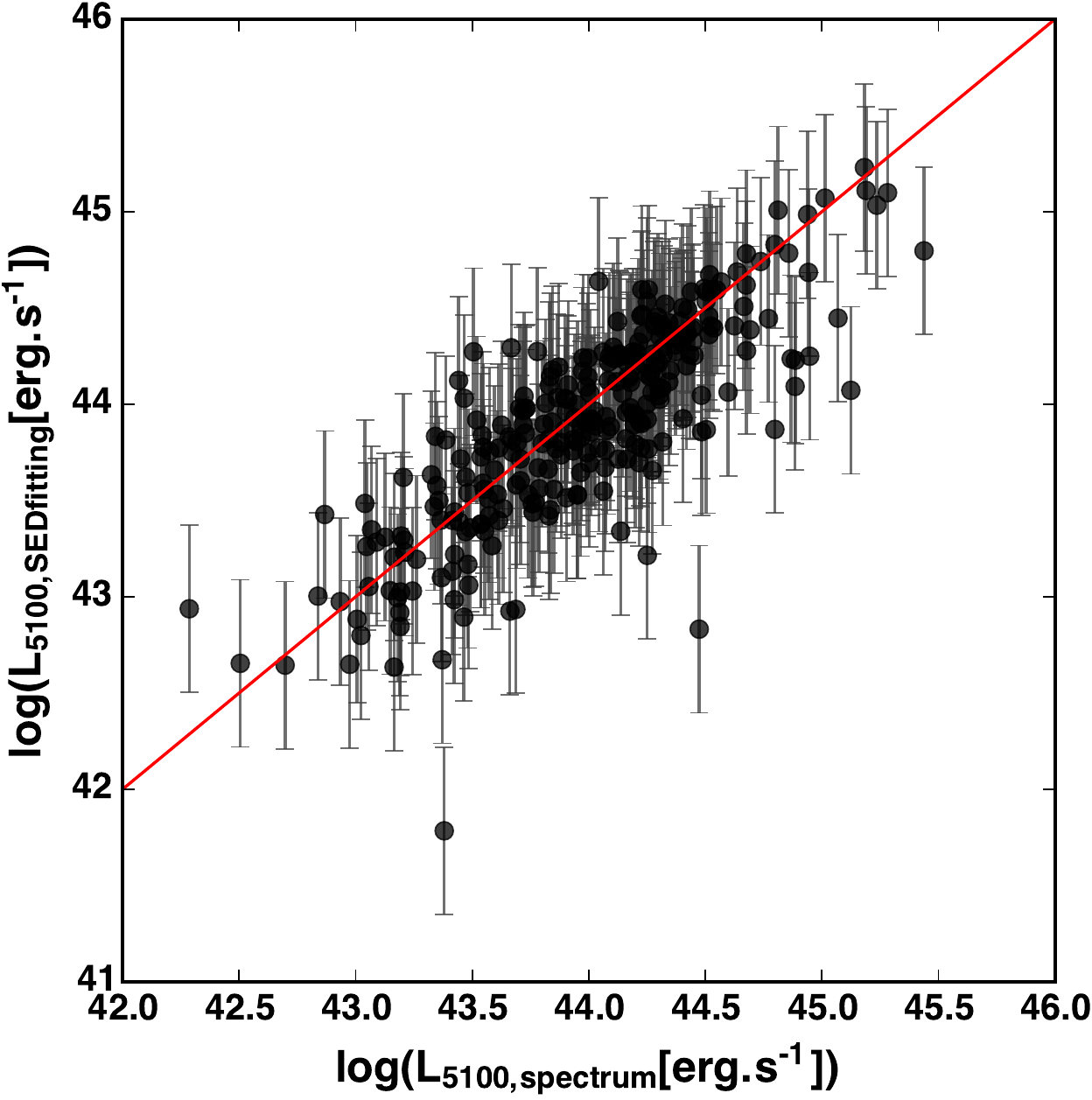}
\epsscale{3.}
\caption{Comparison of SED fitting measurement of $L_{5100}$ with
  measurement from spectra. We compare the estimate of $L_{5100}$
  obtained from \pc (y-axis) with the one obtained directly from the
  spectrum (x-axis). \label{fig_l5100_check}}
\end{figure}

One of the main AGN properties we will use hereafter is the luminosity at
5100\,\AA, $L_{5100}$. Using a spectroscopic sample, we check how well
we can recover $L_{5100}$. We use all the objects in our sample with a
counterpart in SDSS DR12 having a spectrum classified as a
quasar. We correct the spectra for Galactic extinction using the
\citet{Schlegel_1998} map, and the extinction curve from
\citet{Cardelli_1989}, with $R_{V} = 3.1$; we also shift the spectra
to restframe using the listed redshift. We are only interested here in
the continuum luminosity, so we fit the continuum with a power law
($L_{\lambda}=A\lambda^{\alpha}$) for $2500<\lambda [\AA]<5500$, and
exluding the regions of the spectra around the Mg\,II, H$\beta$, and
O\,III lines. We then obtain the luminosity at 5100\,\AA~as $\lambda
L_{\lambda}$. This measure can be contaminated by the host luminosity,
so we use eq.~1 from \citet{Shen_2011} to correct our estimate from
host contribution. On the other hand, we fit the broadband photometry
SED of the same objects with \pc. We compare the luminosities at
5100\,\AA~we derive from the SED and the spectra in
Fig. \ref{fig_l5100_check}.

In Fig. \ref{fig_l5100_check}, we show the results of the SED fitting
using AGN of Type 1, as among our objects with spectroscopic redshifts and
classifications, only 5\% are Type 2. As can be seen in Fig. \ref{fig_l5100_check},
there is excellent agreement between the two methods.

\section{Results}\label{sec_results}

\subsection{Stellar Mass Distribution}\label{sec_mstar}

\begin{figure}
\includegraphics[width=\hsize]{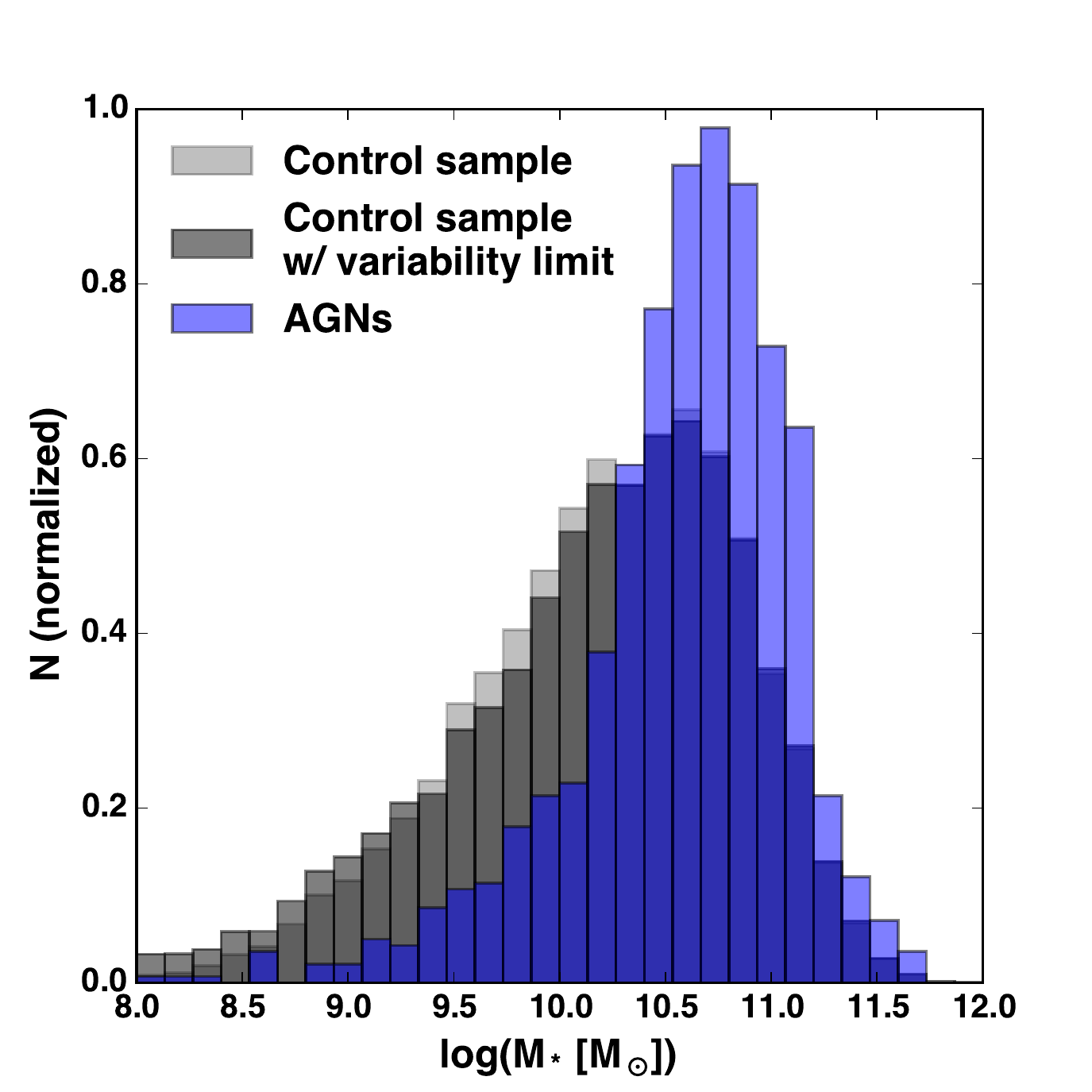}
\caption{Stellar mass distribution. The blue shaded histogram shows
  the stellar mass distribution of the AGN sample. The gray shaded
  histogram shows the distribution of the full control sample, and the black shaded histogram the distribution of the control sample with an additional cut imposed on simulated variability. \label{fig_st_mass}}
\end{figure}

In Fig. \ref{fig_st_mass} we show the stellar mass distribution of the
AGN derived by \pc~(blue shaded histogram). We compare it with that of
the control sample (gray shaded histogram). Fig \ref{fig_st_mass}
suggests that the stellar mass distribution of AGNs is different from
inactive galaxies: it is skewed towards large masses ($\sim
10^{10.6}M_{\odot}$). We checked whether the variability selection can
explain this trend, by simulating the level of variability for
galaxies in our control sample. We assume that each galaxy in the
control sample hosts an AGN, and assign a black hole mass $M_{\rm BH}$
using the total stellar mass - black hole mass relation from
\citet{Bennert_2011}. We then convert $M_{\rm BH}$ to Eddington
luminosity, and further to a bolometric luminosity using the Eddington
ratio distribution from \citet{Kelly_2010}. Using the bolometric
correction of \citet{Krawczyk_2013}, we obtain an estimate of
$L_{5100}$. We assume that the AGN have a power law SED
$f_{\lambda}\propto \lambda^{-1.5}$, and use this luminosity as
normalization. We derive the AGN flux in the PS1 $g$ filter, and
finally use the relation we observe between the fractional variability
$\Delta f/f$ and $L_{5100}$ (see Sect. \ref{sec_var}). We compute what
the brightest magnitude $min_{\rm g}$ is for this simulated variable
source, and keep only objects with $min_{\rm g}<23$\,mag (the
sentivity limit for the detection of sources in the PS1 difference
imaging). The resulting stellar mass distribution is shown in shaded
black in Fig. \ref{fig_st_mass}. It is clear that the variability cut
has little impact on the control sample stellar mass
distribution. This is expected as the AGN luminosity is correlated
with the host stellar mass \citep[given the black hole
    mass-stellar mass correlation, e.g.][]{Haring_2004, Kormendy_2013,
    McConnell_2013, Reines_2015}, but anti-correlated with the
amplitude of variability.

Thus the results presented in Fig. \ref{fig_st_mass} suggest that AGN
are mostly hosted by larger stellar mass galaxies than the underlying
galaxy population.


\subsection{Color Distribution}
\begin{figure}
  \includegraphics[width=\hsize]{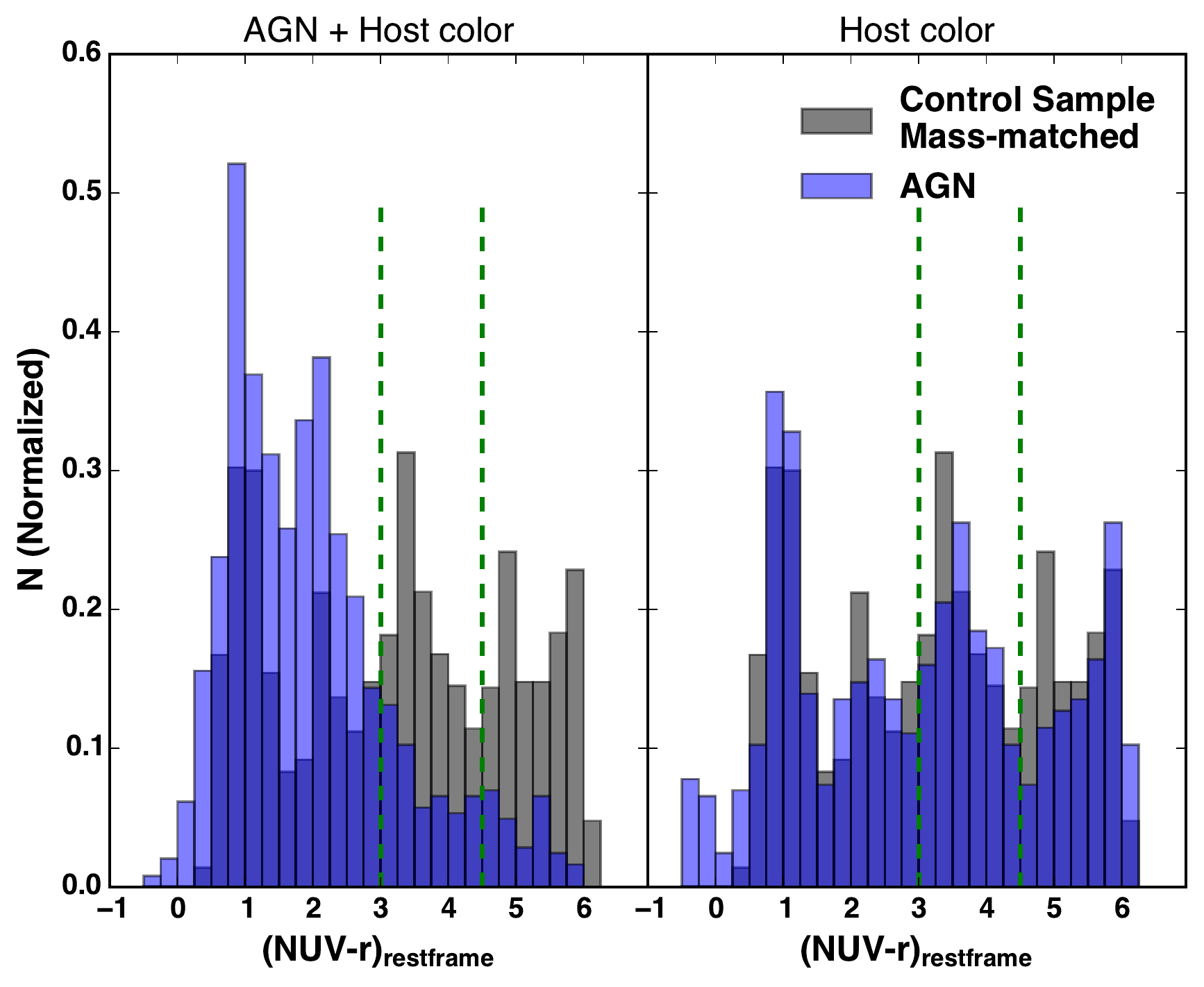}
  \caption{$(NUV-r)_{\rm restframe}$ distribution. The gray shaded
    histograms show the distribution of the mass-matched control
    sample, while the blue shaded histograms show the distribution of
    the AGN sample. Left panel shows the distribution of the total
    $(NUV-r)_{\rm restframe}$ color for the AGN (i.e. including both
    the AGN and the host). Right panel shows the distribution of the
    host color only for the AGN sample. On both panels, the
      vertical dash lines show the limits for the Green
      Valley.\label{fig_nuv_r}}
\end{figure}

We show in Fig. \ref{fig_nuv_r} (left panel) the distribution of the
total restframe $NUV-r$ color, i.e. the combination of the host and
AGN. Note that this color is corrected for dust attenuation. Given the
different mass distributions of the control sample and the AGN, we use
here a mass-matched version of the control sample. We show as
  dashed lines on Fig. \ref{fig_nuv_r} the limits of the Green Valley
  \citep[e.g.][]{Wyder_2007}. As expected, compared to the control
sample, the distribution of AGN host total colors peaks in the blue
sequence, as the AGN emission constributes significantly to the SED,
and can dominate over the stellar populations. In the right panel, we
show the distribution of the host galaxy only restframe $NUV-r$ color,
given by our SED fitting decomposition. Here the distribution of the
host colors is strikingly similar to that of the control sample. In
other words, our results show that hosts of AGN have the same color
distribution than regular, non-AGN galaxies. There is no obvious link
between harboring an AGN and the restrame $NUV-r$ color of its host
galaxy stellar population.

\subsection{Amplitude of variability}\label{sec_var}
\begin{figure*}
  \plottwo{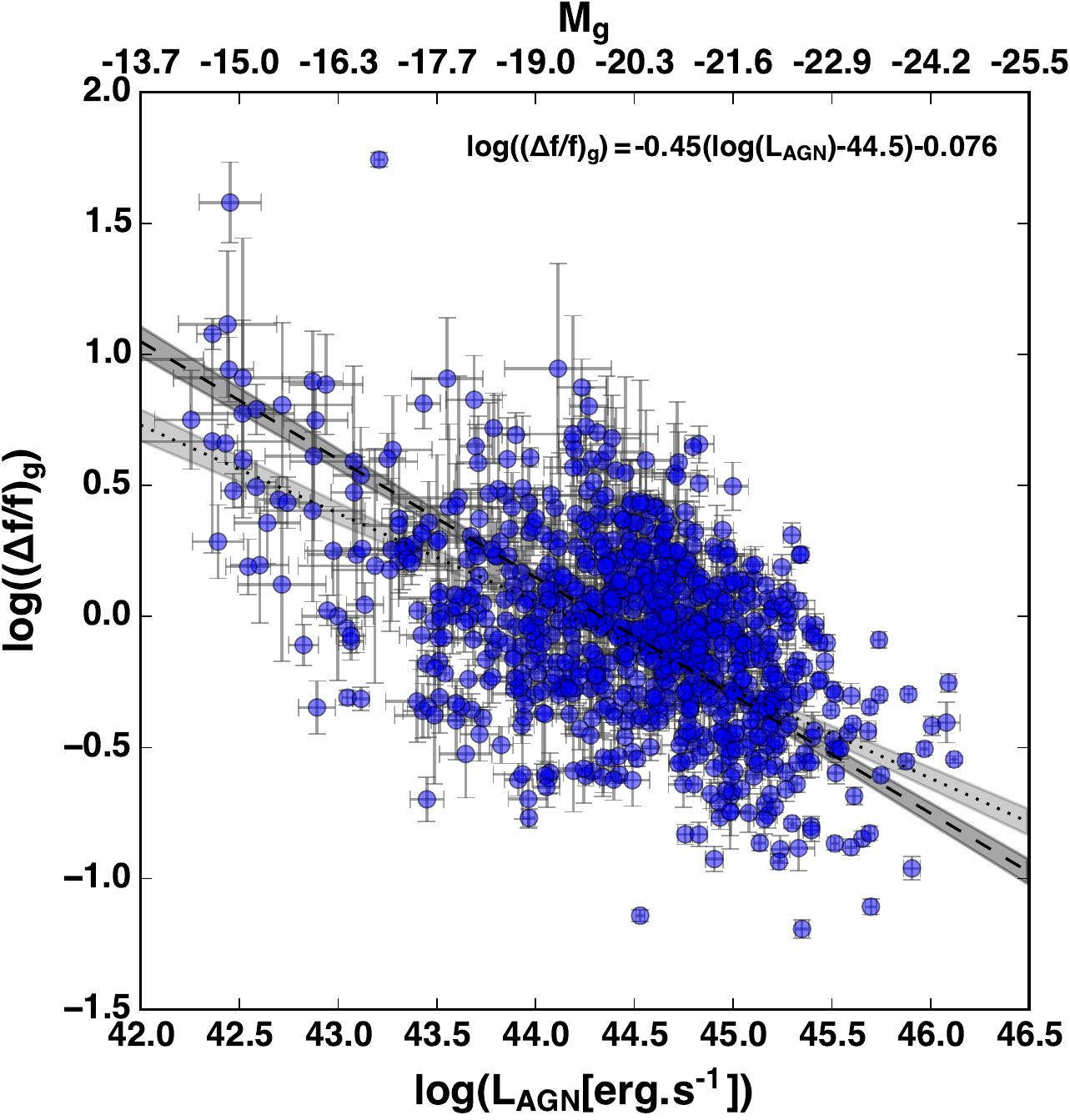}{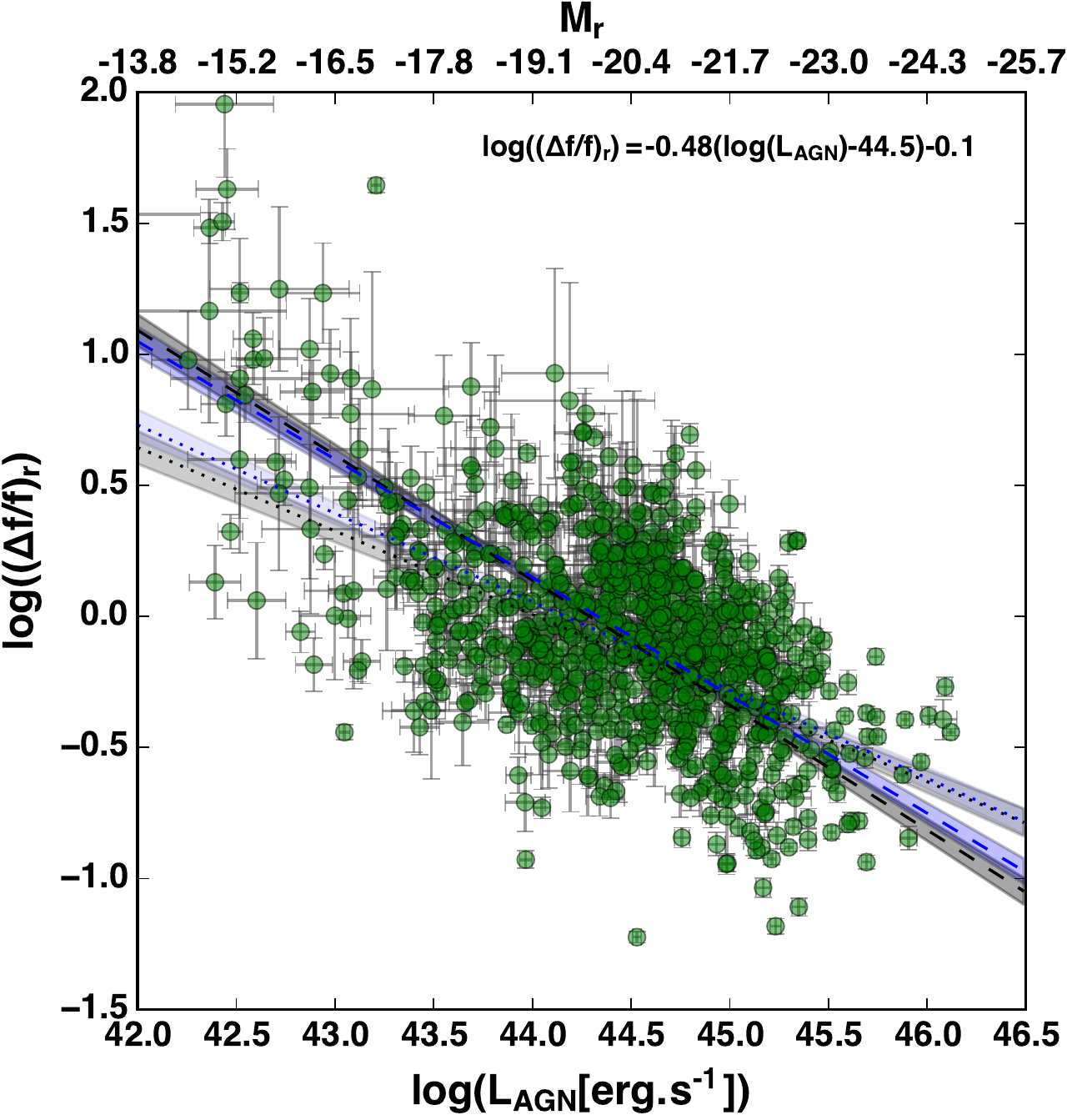}
  \plottwo{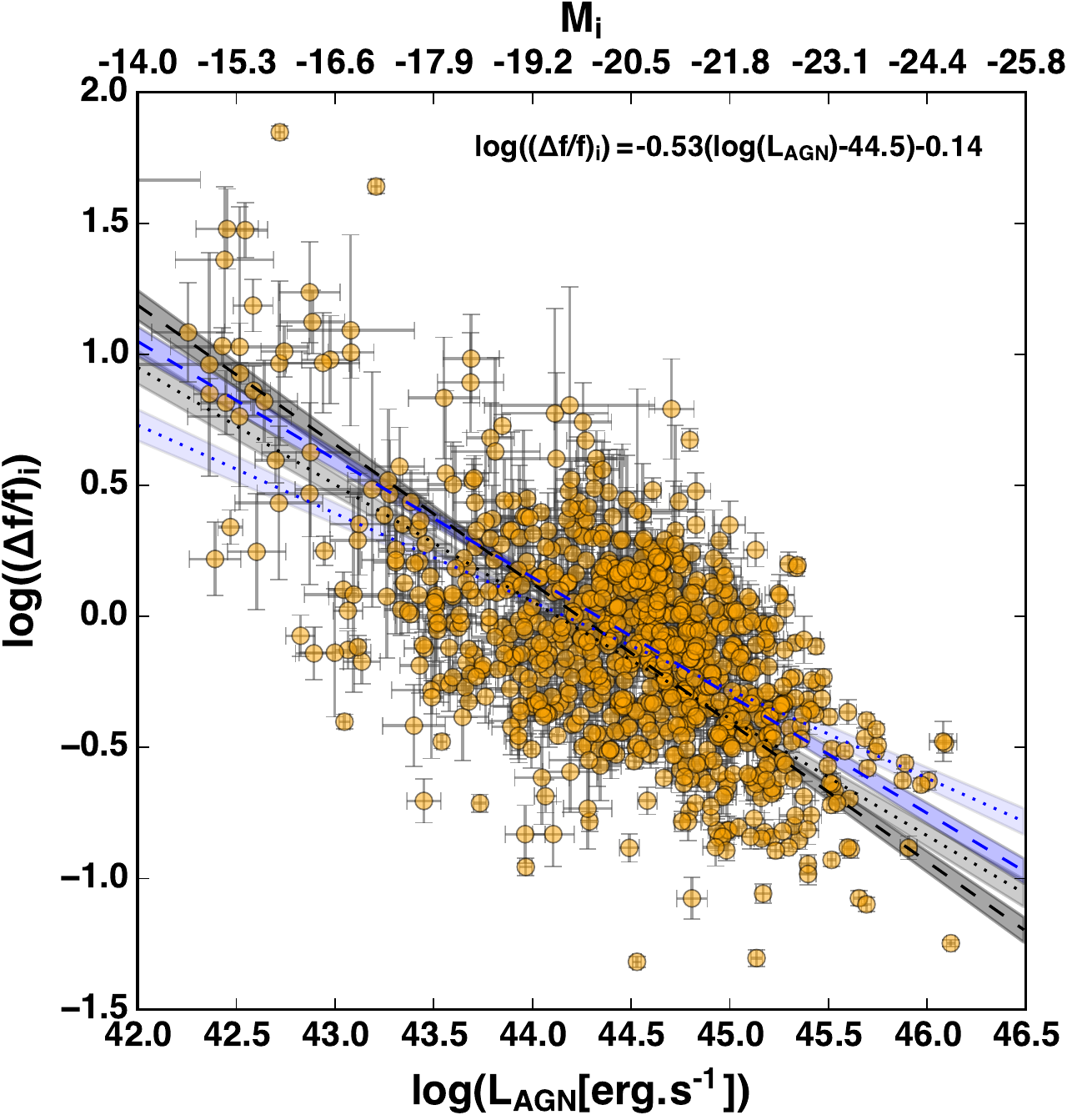}{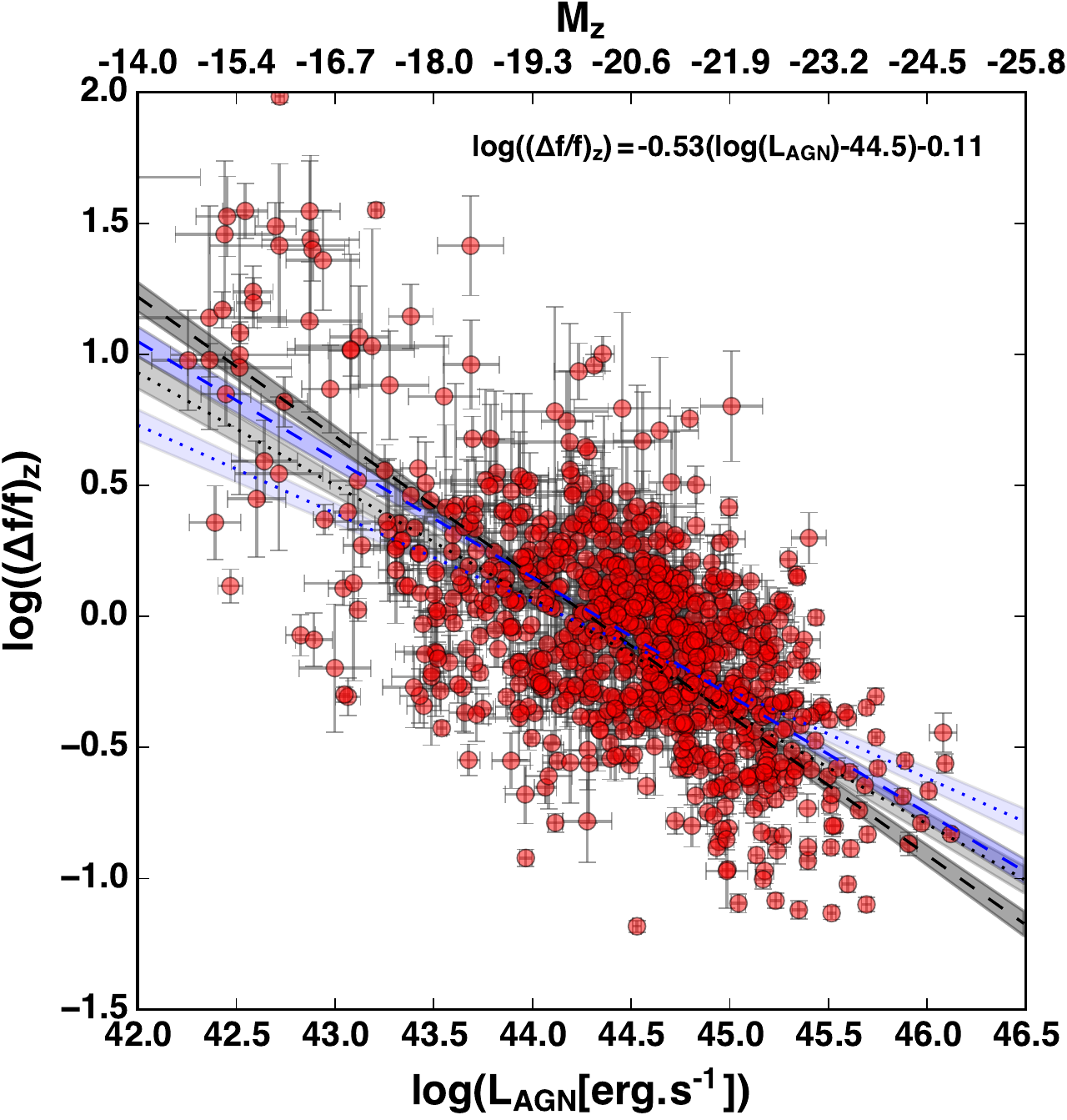}
  \caption{Relative variability amplitude in the $g,r,i,$ and $z$ bands as a function of AGN bolometric luminosity. The upper x-axis shows the corresponding AGN absolute magnitude. In each panel, we show the best fit to eq. \ref{eq_df_f} as a black dashed line, and the dashed area around it represents the errors on the fit. Similarly, we also show the fit to eq. \ref{eq_df_f} using only objects with $L_{\rm AGN}>10^{43.5}\,\rm{erg.s}^{-1}$ as a dotted line. On the panels representing the results for the $r, i, $ and $z$ bands, we show in blue the fits obtained in the $g$ band as reference.\label{fig_df_f}}
\end{figure*}

We investigate the relation between the amplitude of the AGN variability and the AGN luminosity. We use the measure from the image differencing of the minimum magnitude in the $g,r,i,$ and $z$ bands, i.e. the maximum flux of the variable component of the AGN measured over the course of the survey. We convert this magnitude to an absolute magnitude, e.g. $M_{g, \rm min}$ and derive the relative amplitude of the variability as

\begin{equation}
\log\left(\frac{\Delta f}{f}_{\rm g}\right) = -0.4(M_{g, \rm min} - M_g)
\end{equation} 

where $M_g$ is the $g$-band absolute magnitude of the AGN component derived from the SED fitting, and proceed similarly for the other bands. We derive the AGN bolometric luminosity, $L_{\rm AGN}$, using the relation from \citet{Netzer_2014} for the bolometric correction at 5100\,\AA: $b_{5100}=53-\log{L_{5100}}$.                          

We show in Fig. \ref{fig_df_f} the relation between this measure of variability in the $g,r,i,z$ bands and $L_{\rm AGN}$. We use here only objects (831) that have a measure of the minimum apparent magnitude in $g,r,i,z$ bands.
We adjust the relations between $\Delta f/f$ and $L_{\rm AGN}$ by:

\begin{equation}\label{eq_df_f}
  \log\left(\frac{\Delta f}{f}\right) = \beta\left[\log(L_{\rm AGN})-44.5\right] +\log\left(\frac{\Delta f}{f}\right)_0
\end{equation}

We use Orthogonal Distance Regression to perform these fits. This
method allows to take into account errors both on $\frac{\Delta f}{f}$
and $L_{\rm AGN}$. The errors on the parameters for these fits that we
quote are the standard errors on the estimated parameters, which are
derived from the covariance matrix estimated during the fit. We list
in table \ref{tab_var_fit} the best fit parameters in the $g,r,i,$ and
$z$ bands.

\begin{deluxetable}{ccccc}
  \tablecaption{Best fit parameter values for eq. \ref{eq_df_f} in the $g,r,i,$ and $z$ bands.\label{tab_var_fit}}
\tablehead{\colhead{} & \multicolumn{2}{c}{All objects} & \multicolumn{2}{c}{$L_{\rm AGN} > 10^{43.5}\rm{erg.s}^{-1}$}\\[0.1cm]
\colhead{Band} & \colhead{$\beta$} & \colhead{$\log\left(\frac{\Delta f}{f}\right)_0$} & \colhead{$\beta$} & \colhead{$\log\left(\frac{\Delta f}{f}\right)_0$}}
  \startdata
  $g$ & $-0.45\pm0.02$ & $-0.08\pm0.01$ & $-0.34\pm0.02$ & $-0.11\pm 0.01$ \\
  $r$ & $-0.48\pm0.02$ & $-0.10\pm0.01$ & $-0.32\pm0.02$ & $-0.15\pm 0.01$\\
  $i$ & $-0.53\pm0.02$ & $-0.14\pm0.01$ & $-0.45\pm0.02$ & $-0.17\pm 0.01$\\
  $z$ & $-0.53\pm0.02$ & $-0.11\pm0.01$ & $-0.43\pm0.02$ & $-0.15\pm 0.01$
  \enddata
\end{deluxetable}

We note that our sensitivity to AGN variability is set by our
magnitude limit in the difference images of $m_{\rm lim} \sim 23$ mag.
Thus, for a variable AGN, a bright AGN will be detected at smaller
fractional variability ($\Delta$ f/f) amplitudes than a faint AGN:
i.e., the faint AGN in our sample require larger $\Delta f/f$ than the
bright AGN in order to be detected.  We address this bias directly in
our simulation described in Section \ref{sec_mstar} and find that this
does NOT bias us against detecting AGNs in low-mass galaxies.

In all bands considered here, the overall trend is that the amplitude
of the variability decreases with AGN luminosity. Moreover, this
relation is steeper at longer wavelengths. Including all objects in
the fit, we find that $\beta$ decreases from $\sim -0.45$ in the $g$
band to $\sim -0.53$ in the $z$ band. This steepening of the relation
is however mostly due to the objects fainter than $L_{\rm AGN}
<10^{43.5} \rm{erg.s}^{-1}$. We also perform the fit excluding these
objects, and find that while the decrease in $\beta$ is less pronounced
($-0.34$ in $g$ to $-0.43$ in $z$), it is still significant.

The trends we observe that variability amplitude decreases with AGN
luminosity has been noted by a number of studies
\citep[e.g.][]{Hook_1994, Trevese_1994, VandenBerk_2004, Wilhite_2008,
  Bauer_2009, Zuo_2012, GAS_2014}. This trend suggests that AGN
variability can be interpreted by Poissonian models. The slope of the
variability-luminosity relation $\delta L/L\propto L^{\beta}$ is
expected to be $\beta =-0.5$ in that case. When we consider all objects in the fit,
we find slopes consistent with Poissonian models in $r, i, $ and $z$ bands, while the slope is shallower in $g$. When we consider only objects with 
$L_{\rm AGN} >10^{43.5} \rm{erg.s}^{-1}$, we find in all bands                   
shallower slopes than $\beta =-0.5$.

\begin{figure}
  \includegraphics[width=\hsize]{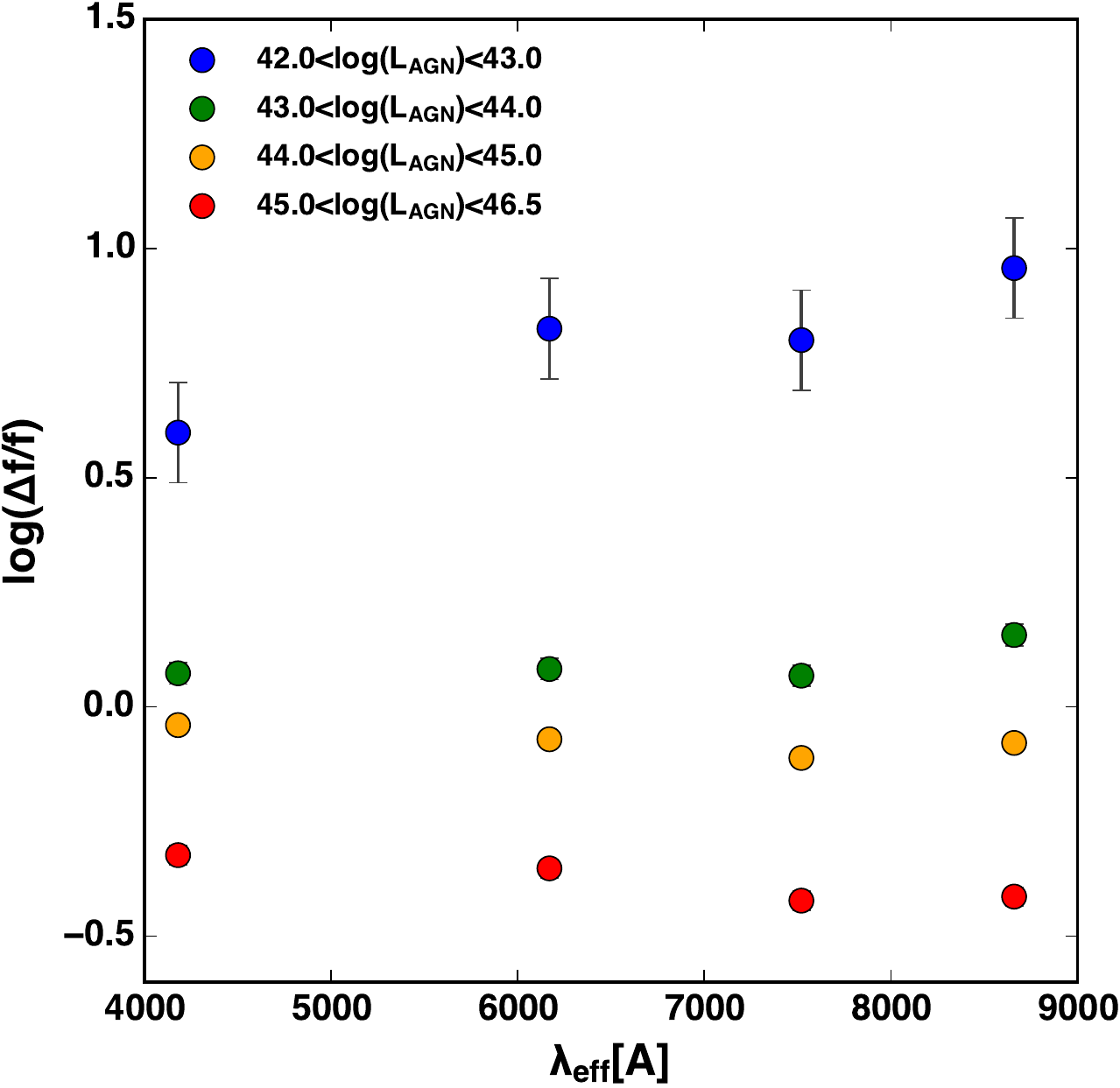}
  \caption{Average relative variability amplitude as a function of
    restframe wavelength in four bins of AGN
    luminosity.\label{fig_df_f_lambda}}
\end{figure}

\citet{GAS_2014} also observed that the slope of the variability
function steepens between the $g$, $r$, and $i$ SDSS bands for Type 1
AGN. We note that \citet{GAS_2014} constrained the variability
function down to $M_i \sim -18.5$, while we extend here the range of
measurements down to $M_i \sim -14.$ Numerous studies
\citep{VandenBerk_2004,Zuo_2012,GAS_2014} have reported that the
amplitude of variability is larger at bluer wavelengths. Thanks to our
sample spanning a larger range of bolometric luminosities, we can
revisit this claim. At $L_{\rm AGN} \gtrsim 10^{43.5}
\rm{erg.s}^{-1}$, we observe that on average the amplitude of
variability is larger in $g$ than in the other bands, which is
consistent with previous results. At fainter luminosities however,
which were not sampled by previous studies, AGN display larger
variability amplitudes in redder bands.  We examine in
Fig. \ref{fig_df_f_lambda} the wavelength and luminosity dependence of
the relative variability amplitude. We show in four bins of
$\log(L_{\rm AGN})$ the average relative variability amplitude as a
function of restrame wavelength. The errors bars are derived by
propagating the errors in the mean. These measures show that for
$43<\log(L_{\rm AGN}\rm{[erg.s}^{-1}\rm{]})<45$, there is no
significant wavelength dependence of the AGN variability (i.e. the
relation is consistent with a flat one at the $1\,\sigma$ level). For
bright AGNs, at $45<\log(L_{\rm AGN}\rm{[erg.s}^{-1}\rm{]})<46.5$, we
observe that the variability decreases at redder wavelengths
($2.5\,\sigma$ level), while for faint AGNs, at $42<\log(L_{\rm
  AGN})\rm{[erg.s}^{-1}\rm{]}<43$, the trend reverses, as the
variability increases at redder wavelengths ($2.7\,\sigma$ level).

\subsection{AGN and host SFR}
\begin{figure}
  \includegraphics[width=\hsize]{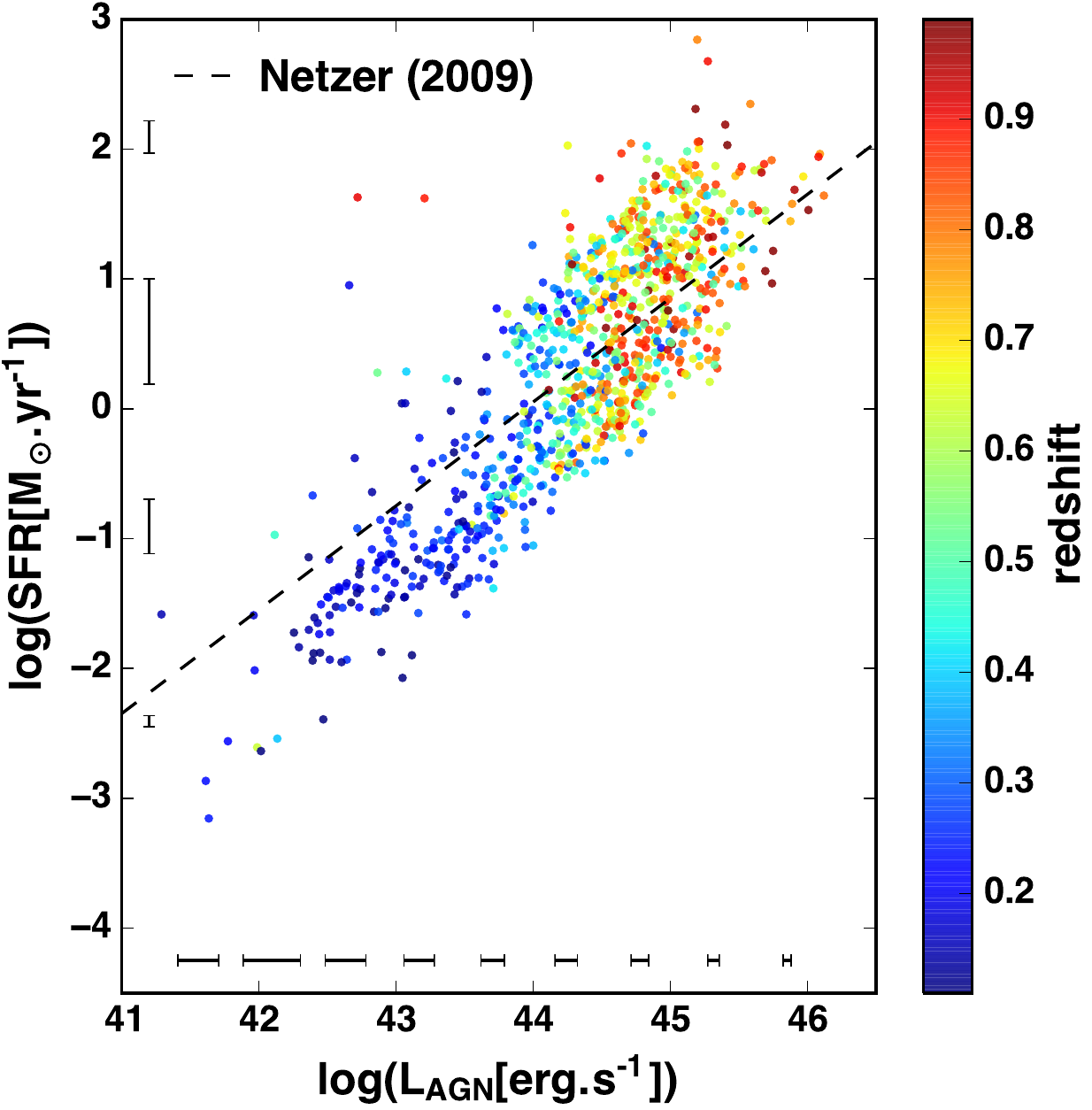}
  \caption{SFR-bolometric AGN luminosity. Points are color coded by
    the redshifts. Error bars show the median error in bins of SFR or
    $L_{\rm AGN}$.\label{fig_sfr_lagn}.}
\end{figure}

We investigate in Fig. \ref{fig_sfr_lagn} the relation between the
host SFR and the AGN bolometric luminosity.  Our results show that for
variability selected AGN, there is an overall good correlation between
SFR and the bolometric AGN luminosity. We also color code in
Fig. \ref{fig_sfr_lagn} the symbols by the redshifts of the
objects. This shows that higher redshift objects display higher SFRs
and bolometric AGN luminosity.  The observation that the commonly
observed SFR-$L_{\rm AGN}$ relation is built from the contribution of
galaxies at various redshifts is consistent with the results from
\citet{Stanley_2015} based on a X-ray selected
sample. \citet{Stanley_2015} furthermore found that, at a given
redshift, there is no correlation between SFR and $L_{\rm AGN}$: At a
given redshift ($0.2<z<2.5$), the relation between SFR and $L_{\rm
  AGN}$ is mostly flat, while the amplitude of this relation increases
with redshift. The superposition of these relations yields the
impression of an overall correlation between SFR and $L_{\rm AGN}$. In
contrary to \citet{Stanley_2015}, the overall relation between SFR and
$L_{\rm AGN}$ we observe is preserved at all redshifts sampled
here. We note that \citet{Stanley_2015} use a different technique to
determine the SFR, which is likely to yields overestimates in the case
of Type 2 AGNs \citep{Ciesla_2015}. We compare our results with the
average SFR-$L_{\rm AGN}$ relation observed by \citet[][dashed line on
  Fig. \ref{fig_sfr_lagn}]{Netzer_2009} from a sample at low redshift
($z\sim 0.1$) of Type I and Type II AGN. \citet{Netzer_2009} further
shows that higher redshift ($z\sim 2-3$) QSOs from \citet{Lutz_2008}
also fall on this relation, at higher SFRs and $L_{\rm AGN}$.  Our
results are in excellent agreement with the relation from
\citet{Netzer_2009}. We note that the object selection used by
\citet{Netzer_2009} is different from the one we use as it combines
Types I and II selected from spectral features in the restframe
optical; moreover the methods \citet{Netzer_2009} used to derive SFR
and $L_{\rm AGN}$ are completely different from ours.\\ We also
checked that our SED fitting technique enables to probe the whole
range of SFR and $L_{\rm AGN}$. In details, the models we use do probe
the whole range; moreover the actual values of the parameters (SFR and
$L_{\rm AGN}$) we use are derived from the PDF built during the SED
fitting which also allows a larger spread around the models.\\ In a
recent work, \citet{Rosario_2012} studied the properties of X-ray
selected AGN at $z<2.5$ using constraints in the FIR from
Herschel/PACS data. They found that at high $L_{\rm AGN}$
luminosities, the SFR-$L_{\rm AGN}$ relation follows a trend similar
to the one we observe, but at lower luminosities, the average SFR is
constant. Their findings are in line with the earlier results from
\citet{Lutz_2010} who performed stacking at 870\,$\mu$m using similar
AGN samples. These two regimes in the SFR-$L_{\rm AGN}$ relation are
expected to reflect the two regimes of blak hole growth,
starburst-like at high $L_{\rm AGN}$, and "hot halo'' at low $L_{\rm
  AGN}$ \citep[e.g.][]{Gutcke_2015}. Our results do not support these
observations. We note that \citet{Rosario_2012} results are based on
the luminosities at 60\,$\mu$m, while we derive here a SFR from a full
SED modelling. \citet{Rosario_2012} argue that their 60\,$\mu$m
luminosity estimates are not contaminated by AGN emission. As
mentioned by \citet{Rosario_2012}, AGN contribution to the SED at
60\,$\mu$m would require large dust torii \citep{Fritz_2006}, which
are thought to be rare.\\ A potential reason for the difference
between our results and those from \citet{Rosario_2012} is that our
selection does not bias against quiescent galaxies, or galaxies with
very low star formation rates. Note that \citet{Salvato_2009} showed
that the fraction of quiescient galaxies in the XMM-Newton COSMOS
sample (the catalog used by \citet{Rosario_2012} in the COSMOS field)
is small. Another possible reason for the discrepancy is that we do
not probe well AGN fainter than $L_{\rm AGN} \lesssim
10^{42.5}\rm{erg.s}^{-1}$, a range where \citet{Rosario_2012} observe
the flattening of the SFR-$L_{\rm AGN}$ relation. Our variability
selection is also biased against fainter sources.  Due to photometric
errors, faint AGN require a larger fractional variability to be
detected than bright AGN (see Sect.  \ref{sec_var}).  Thus, this will
translate to a luminosity threshold at a given redshift range below
which we are unable to detect variability.

\subsection{AGN and the Main Sequence of Star Formation}
\begin{figure*}
  \includegraphics[width=\hsize]{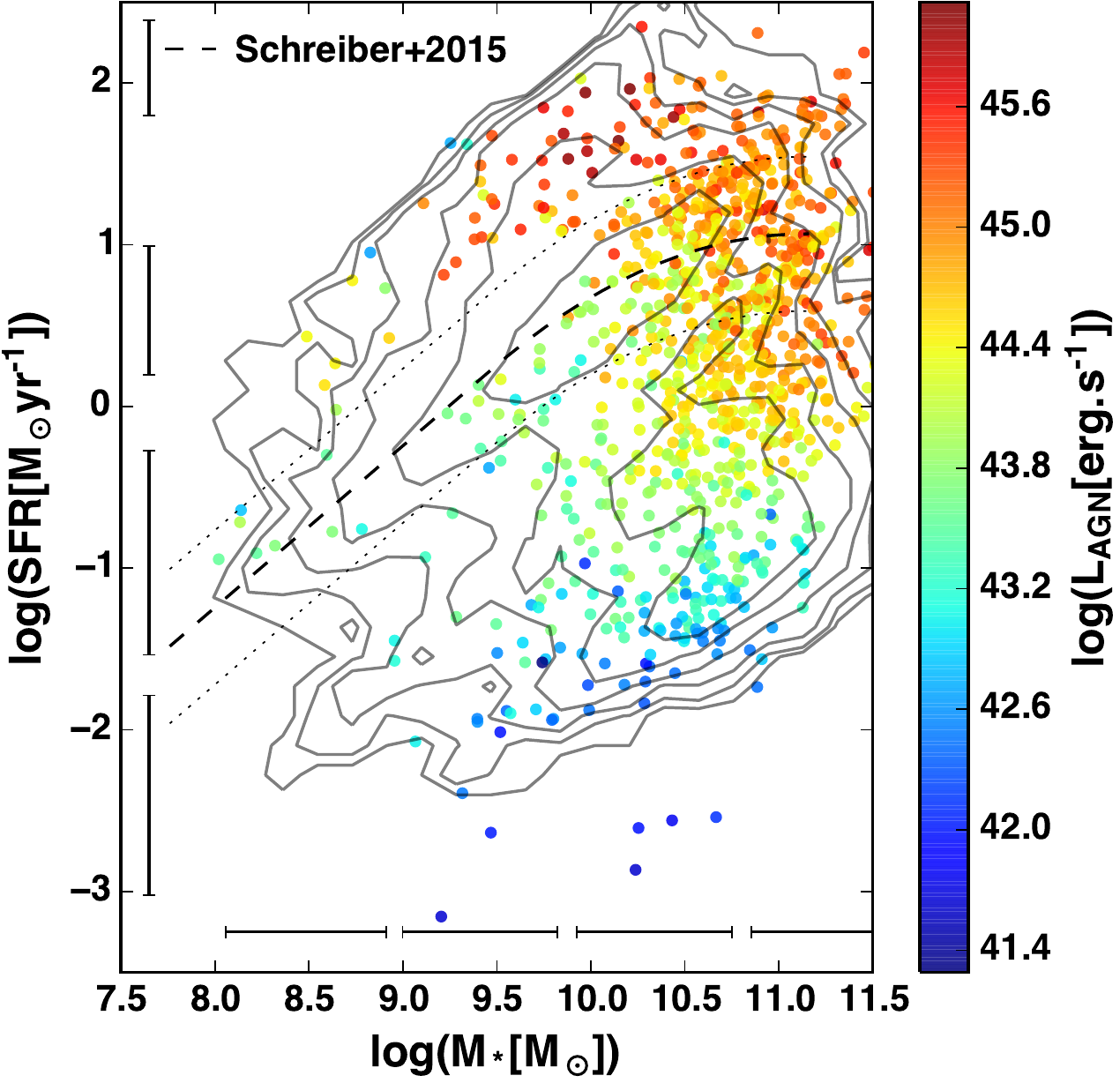}
  \caption{SFR-$M_*$ relation. The gray contours show the relation for
    the control sample. The filled circles represent the AGN sample,
    color coded by $L_{\rm AGN}$. Error bars show the median error in bins of SFR or
    $M_*$. The dashed line shows the fit to the main sequence of star forming galaxies from \citet{Schreiber_2015} at the median redshift of the sample. The dotted lines shows the limits we use for the main sequence, assuming a dispersion of 0.3\,dex.
 \label{fig_sfr_m}}
\end{figure*}

In Fig. \ref{fig_sfr_m} we show the location of AGN in the SFR-$M_*$
plane, color-coded by the AGN bolometric luminosity, along with the
distribution of galaxies in the control sample as contours. We also
show the fit to the locus of the main sequence of star-forming
galaxies obtained by \citet{Schreiber_2015}, at the median reshifts of
the control sample ($z_{\rm median}= 0.39$), after converting their
results to a Chabrier IMF.

\begin{figure}
  \includegraphics[width=\hsize]{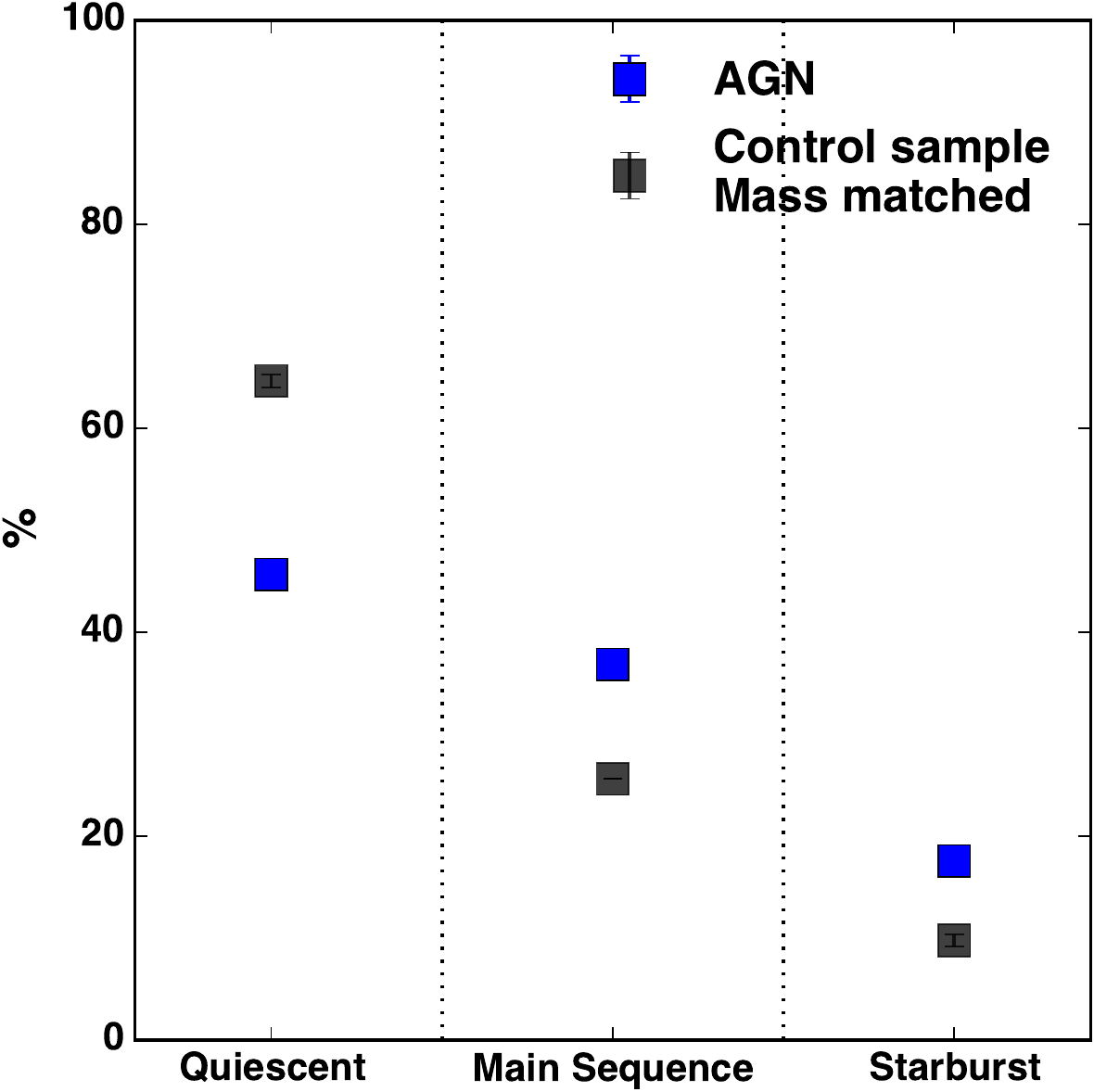}
  \caption{Percentage of AGN hosts (blue squares) below the main sequence, ('Quiescent'), on the main sequence, and above ('Starburst'). Percentages for the mass matched control sample are shown in black squares. Error bars on percentages are displayed, but are smaller than the size of the symbols.
 \label{fig_ms_fractions}}
\end{figure}

A number of studies have extensively investigated the relation between
SFR and $M_*$ for inactive galaxies\,, from low to high redshifts
\citep[e.g.][]{Elbaz_2007,Noeske_2007,Karim_2011,Heinis_2014,Schreiber_2015}. Our
control sample clearly displays the star-forming sequence with a
relation between SFR and $M_*$ similar to the results obtained by
\citet{Schreiber_2015}, while the bulk of quiescent galaxies are
located at lower SFRs and relatively high $M_*$ ($\sim
10^{10.75}M_{\odot}$).  Above a stellar mass of $\sim 10^{9.5}
M_{\odot}$, AGN occupy the full range of SFRs. The bolometric AGN
luminosity is mostly correlated with SFR, with a weaker additional
correlation with $M_{*}$. \\ Running a principal component analysis (PCA)
on SFR, $L_{\rm AGN}$, and $M_*$ shows that in this space, $M_*$
encodes only 5\% of the available information. While the distributions
of AGN hosts and inactive galaxies look similar above $\sim 10^{9.5}
M_{\odot}$, they are not strictly-speaking drawn from the same parent
distribution, according to Kolmogorov-Smirnov and Mann-Whitney
statistics. There is for instance virtually no AGN at $M_* >
10^{11.25}M_{\odot}$ and SFR$<1 M_{\odot}\rm{yr}^{-1}$. \\We do observe
however a significant fraction of AGN hosts within the main sequence of
star forming galaxies. Assuming that the dispersion around the main
sequence is 0.3 dex, we determine the percentages
of AGN hosts that are above the main sequence, within the main
sequence, or below the main sequence. We also perform the same for our
control sample, mass matched this time.  We derive errors on these
fractions using the errors on the stellar mass and SFRs. The results
are shown in Fig. \ref{fig_ms_fractions}. The percentages of AGN hosts
and control sample galaxies show the same trend, that they decrease
from quiescent to starburst. However, the percentage of AGN in
quiescent hosts is significantly lower than for the control sample. On
the other hand, the occurrence of AGN in MS or starburst hosts is
larger than for the control sample. These results suggest that AGN
activity is, at least moderately, linked to star formation activity.

\section{Discussion}\label{sec_discussion}

\subsection{AGN variability}
We first note that our variable AGN sample has been selected from the
PS1 transient alerts, which are by definition variable at the
$5\,\sigma$ level in at least 3 epochs in a time window of 15 days,
properly measuring signal-to-noise in the individual difference images
(see Sect. \ref{sec_var_sel_agn}).  This selection criteria does mean
that fainter sources must have a larger fractional variability than
brighter sources (see Sect.  \ref{sec_var}) to be detected at the
$5\,\sigma$ level.  However, we find in our simulations that
regardless of this bias, we are still capable of detecting AGN across
the full range of host galaxy masses (see Fig. \ref{fig_st_mass}).
Similarly, the trends shown in Fig. \ref{fig_df_f} and
Fig. \ref{fig_df_f_lambda} cannot be attributed to this selection
bias, since our sensitivity to variability at the low-luminosity end
is well below the observed relations.

Our results on AGN variability are consistent with previous studies
that showed that the variability amplitude decreases with AGN
luminosity \citep{Hook_1994,Trevese_1994, VandenBerk_2004,
  Wilhite_2008, Bauer_2009, Zuo_2012, GAS_2014}. This trend has been
interpreted in the context of Poissonian models, where variations are
due to the stochastic superposition of independent flares
\citep[e.g.][]{Cid_Fernandes_2000}. These models predict that the
relation between variability amplitude and luminosity have a slope of
$-0.5$. We do observe slopes consistent with this value, when we
consider all objects in our sample, in the $r, i,$ and $z$ bands. The
fact that we do not observe relations consistent with the Poissonian
case does not actually rule it out. Indeed, selection effects might
yield a not well defined variability relation, and moreover, a slope
of $-0.5$~is only expected in the case of the simplest Poissonian
models, where all the components of the models are universal constants
among all objects \citep{Cid_Fernandes_1996}.

As previously noticed by \citet{GAS_2014}, we observe that the
relation between the variability and $L_{\rm AGN}$ steepens at redder
wavelengths. A potential interpretation for this effect in a
Poissonian context is that there is an underlying nonvariable
background component redder than the SED of the flares
\citep{Cid_Fernandes_2000}.  We find however that this trend is
luminosity dependent: for AGN with $L_{\rm AGN} \gtrsim
10^{43.5}\,\rm{erg.s}^{-1}$ our results show that the variability is
larger in $g$, but for fainter AGN, the trend is reversed. In the
context of accretion disk models \citep{Shakura_1976}, where the
variability is caused by a change in the accretion rate, one expects
the variability to decrease monotonically with wavelength
\citep[e.g.][]{Li_2008}, which is inconsistent with our results at the
faint end.  While the ``bluer when brighter'' AGN are the most
commonly observed, some ``redder when brighter'' AGN have also been
noticed. In particular, \citet{Gu_2006} noticed using data spanning
three months two flat spectrum radio quasars that become redder when
brighter, which was confirmed by \citet{Rani_2010}. \citet{Gu_2006}
interpret this observation as the non-thermal component dominating the
UV-optical region of the spectrum when the source brightens.

\subsection{AGN and host properties}

The link between AGN and their host properties has been widely studied over the last decades, using various techniques and selections. We revisit here this topic thanks to our variability selected AGN. 
\subsubsection{Color distribution}
Our first result is that, after removing the contribution af the AGN
to the observed color, we find that the color distribution of AGN
hosts is similar to that of inactive galaxies. We note here that the
reason the color distribution of our control sample is not bimodal
comes from the fact that we are matching the mass and the i-band
apparent magnitude distributions of the control galaxies to that of
the AGNs. Our result that the color distribution of AGN hosts is
similar to that of inactive galaxies is in contrast with a number of
studies which observed that AGN are particularly common in green
valley galaxies, suggesting that they are responsible for the
quenching of star formation in this population: \citet{Martin_2007}
were one the first authors to make this observation; we however note
that their $NUV-r$ was not corrected from AGN
contamination. \citet{Schawinski_2009} studied the host properties of
a sample of X-ray selected AGNs, and subtracted a central point source
in optical imaging to derive the host optical colors. They claimed
that AGN are mostly found in green valley galaxies. We argue here that
our results are actually consistent with theirs, once selection
effects are taken into account. First of all, we note that
\citet{Schawinski_2009} observe few AGN in hosts fainter than
$M_{\rm{r}}=-20.5$. This is in line with the fact that we observe very
few AGN in host with stellar masses smaller than $M_* \sim
10^{9.5}\,M_{\odot}$. Within the range of $M_{\rm{r}}$ where their
sample probes hosts bluer than the red sequence, the color
distribution of AGN hosts is rather similar to what we
observe. Moreover, the sample of \citet{Schawinski_2009} is limited at
$L_{0.1-2.4\rm{keV}}>10^{42}\,\rm{erg s}^{-1}$, which corresponds to a
bolometric luminosity of $L_{\rm AGN}\sim 10^{43}\,\rm{erg s}^{-1}$,
assuming a bolometric correction of 10. Given the good correlation we
observe between $L_{\rm{AGN}}$ and SFR (Fig. \ref{fig_sfr_lagn}), this
luminosity cut explains why \citet{Schawinski_2009} do not observe any
AGN within the red sequence. In this context, \citet{Xue_2010} showed
that it is essential to properly take into account selection effects
to discuss the relation between AGN and host properties. They note in
particular that it is critical to use mass-matched samples to compare
the color distributions of AGN hosts and non-AGN galaxies. Once this
is taken into account, they also found that these color distributions
are similar, from $z=3$ to $z=0$. \citet{Aird_2012} also found that
the AGN fraction is only moderately enhanced in galaxies with blue or
green colors.

\subsubsection{AGN-SFR connection}
A large number of studies have focused on the link between AGN and SFR
activity, as AGN feedback has been shown in simulations to be a
promising mechanism to quench star formation in galaxies, and explain
the global trends in galaxy evolution since $z=2$. Our point is not to
contradict the fact that AGN \textit{do} quench star formation in some
galaxies. It is clear that AGN can inject in the intergalactic medium
significant amounts of energy that can prevent further gas from
cooling \citep{Croton_2006,Fabian_2012,Tombesi_2015}. However, the
question is rather whether AGN feedback is statistically the main
process that drives the building of the red sequence of galaxies
observed since $z=2$. Our results show that there is a good
correlation between AGN and SFR over the whole ranges of redshift and
bolometric luminosity we probed here, and hence that it is not obvious
that AGN feedback is the dominant process.  Our results for the
SFR-$L_{\rm AGN}$ correlation are in contrast with the results from
\citet{Lutz_2008} and \citet{Rosario_2012} who, based on X-ray AGN
selected samples and stacking in the FIR, observe a plateau in the
SFR-$L_{\rm AGN}$ relation for faint AGN. This can be interpreted
\citep[e.g.][]{Gutcke_2015} by two modes of AGN accretion. The mode
corresponding to the SFR-$L_{\rm AGN}$ corrrelation is the
``starburst'' regime, where galaxies experience starburst and AGN
activity with high SFR and black hole accretion rates. The other mode,
corresponding to the plateau observed by \citet{Lutz_2008} and
\citet{Rosario_2012}, is the ``hot-halo'' regime, where the growth of
the black holes is linked to the AGN feedback mechanism.\\ The
discrepancy between the results of \citet{Lutz_2008} and
\citet{Rosario_2012} and ours is mitigated by the fact that we do not
probe well the range of luminosities where \citet{Lutz_2008} and
\citet{Rosario_2012} observe the plateau ($L_{\rm AGN} \lesssim
10^{43}\,\rm{erg s}^{-1}$). However, the SFR values for the few
objects we observe in that range are not consistent with a
plateau. One possibility is that our SED fitting procedure would
erroneously mistake moderate levels of star formation for moderate
levels of AGN activity. However, this seems unlikely according to the
work of \citet{Ciesla_2015}, who showed that at low levels of AGN
activity the SED procedure we use properly recovers the SFR. According
to the simulations of \citet{Gutcke_2015}, the plateau in the
SFR-$L_{\rm AGN}$ correlation is created by a mix of galaxies on and
out of the main sequence. We note however that there are some
discrepancies between these simulations and the observations. For
instance the $z\sim0$ results from \citet{Rosario_2012} are based on
the \textit{Swift} BAT AGN sample of \citet{Cusumano_2010}. According
to \citet{Koss_2011}, the minimum stellar mass for the hosts of this
sample is around $M_* \sim 10^{9.8} M_{\odot}$. In this case it is
unlikely, according to the simulations of \citet{Gutcke_2015} that
faint AGN with hosts more massive than this limit would have an
average SFR larger than one (see their Fig. 5).\\ In summary, our
results suggest that for $0<z<1$ there is a good correlation between
AGN host SFR and $L_{\rm AGN}$. This implies that the black hole
accretion rates are also well correlated with SFR. These results are
consistent with the picture that AGN hosts experience secular
evolution, and that black holes are mostly fueled by the same
mechanisms that fuel star formation events. We visually inspected the
objects in our sample with higher SFRs, and did not find any merger or
interaction signatures. This does not mean that mergers or
interactions do not trigger AGN activity, but rather it suggests that
these events are not the main channels for AGN accretion. These
results are in line with the results from \citet{Mullaney_2012b} who
found that the accretion rates of supermassive black holes are well
correlated with SFR from $z=2$ to $z=1$.  We note however that while
our results are in line with previous ones suggesting that there is a
good correlation between SFR and black hole activity, we can not get
more insight from these results about the physical mechanisms at play
in this correlation. Moreover, our sample is biased against Type 2
AGNs. Assuming the scenario of \citet{Hopkins_2008} this means that we
would be missing a fraction of the AGN population at stages where the
star formation and AGN activity are still coexisting. This population
of AGNs are hosted by ULIRGs type of objects, that we would miss with
our UV/optical restframe selections.

\section{Conclusions}\label{sec_conclusions}
We studied the properties of $\sim 1000$ hosts of AGN at $z<1$ selected by variability in optical bands in the \ps~survey. Thanks to extensive wavelength coverage from the UV to the FIR, we performed reliable AGN/host decomposition through SED fitting. Our results can be summarized as follows:

\begin{itemize}
  \item We observe AGN in mostly in massive hosts: $M_* \gtrsim 10^{9.5}\,M_{\odot}$
  \item The relative amplitude of AGN variability decreases with AGN bolometric luminosity. This relation steepens with wavelength (between $g$ and $z$ band), and the steepening is driven by faint AGN $L_{\rm AGN} <10^{43.5} \,\rm{erg s}^{-1}$.
  \item The $NUV-r_{\rm restframe}$ color distribution of AGN hosts is similar to a mass-matched control sample of non-AGN galaxies.
  \item We observe a well defined correlation between $L_{\rm AGN}$ and SFR, valid over the whole redshift range we probe, as well as for $10^{42.5}<L_{\rm AGN} <10^{45.5}\,\rm{erg s}^{-1}$
  \item Above $M_* \gtrsim 10^{9.5}\,M_{\odot}$, AGN are most likely to be hosted by Main Sequence or starburst galaxies than by quiescent ones.
  \item These results suggest that there is no obvious correlation between AGN activity and SFR quenching at $z<1$. This is in line with the results of a number of previous studies; however this study does not enable us to point towards a specific fueling mechanism.

\end{itemize}

\acknowledgements It is a pleasure to thank V\'{e}ronique Buat and Laure
Ciesla for help with the \pc~software. We also thank Richard Mushotzky
for stimulating discussions.

The Pan-STARRS1 Surveys (PS1) have been made possible through
contributions of the Institute for Astronomy, the University of
Hawaii, the Pan-STARRS Project Office, the Max-Planck Society and its
participating institutes, the Max Planck Institute for Astronomy,
Heidelberg and the Max Planck Institute for Extraterrestrial Physics,
Garching, The Johns Hopkins University, Durham University, the
University of Edinburgh, Queen's University Belfast, the
Harvard-Smithsonian Center for Astrophysics, the Las Cumbres
Observatory Global Telescope Network Incorporated, the National
Central University of Taiwan, the Space Telescope Science Institute,
the National Aeronautics and Space Administration under Grant
No. NNX08AR22G issued through the Planetary Science Division of the
NASA Science Mission Directorate, the National Science Foundation
under Grant No. AST-1238877, the University of Maryland, and Eotvos
Lorand University (ELTE) and the Los Alamos National Laboratory.

Funding for the DEEP2 Galaxy Redshift Survey has been provided in part
by NSF grant AST00-71048 and NASA LTSA grant NNG04GC89G.  Funding for
PRIMUS is provided by NSF (AST-0607701, AST-0908246, AST-0908442,
AST-0908354) and NASA (Spitzer-1356708, 08-ADP08-0019,
NNX09AC95G). This paper uses data from the VIMOS Public Extragalactic
Redshift Survey (VIPERS). VIPERS has been performed using the ESO Very
Large Telescope, under the "Large Programme" 182.A-0886. The
participating institutions and funding agencies are listed at
http://vipers.inaf.it

\end{document}